\documentclass{iau_FM}
\usepackage{graphicx}
\usepackage{natbib}
\usepackage{color}

\newcommand{\msun}{\,{\rm M_\odot}}

\newcommand{\beq}{\begin{equation}}
\newcommand{\eeq}{\end{equation}}

\title[FM~14.~~Gravitational Wave Symphony of Structure Formation] 
{Massive black holes in merging galaxies}

\author[Volonteri, Bogdanovi\'c, Dotti \& Colpi]   
{Marta Volonteri$^1$, Tamara Bogdanovi\'c$^2$, Massimo Dotti$^3$ \and Monica Colpi$^3$}

\affiliation{$^1$Institut dÕAstrophysique de Paris, Sorbonne Universit\`{e}s, UPMC Univ Paris 6 et CNRS, UMR 7095, 98 bis bd Arago, 75014 Paris, France \\ email: {\tt martav@iap.fr} \\[\affilskip]
$^2$ Center for Relativistic Astrophysics, School of Physics, Georgia Tech 837 State Street, Atlanta, GA 30332-0430, USA \\ email: {\tt tamarab@gatech.edu} \\[\affilskip]
$^3$ Department of Physics \& INFN, University of Milano Bicocca, Piazza della Scienza 3, I20126 Milano, Italy \\email: {\tt massimo.dotti@mib.infn.it, monica.colpi@mib.infn.it}}

\pubyear{2015}
\jname{Astronomy in Focus, Volume~1} 
\editors{Piero Benvenuti, ed.}
\begin{document}

\maketitle

\begin{abstract}
The dynamics of massive black holes (BHs) in galaxy mergers is a rich field of research that has seen much progress in recent years. In this contribution we briefly review the processes describing the journey of BHs during mergers, from the cosmic context all the way to when BHs coalesce. If two galaxies each hosting a central BH merge, the BHs would be dragged towards the center of the newly formed galaxy. If/when the holes get sufficiently close, they coalesce via the emission of gravitational waves.  How often two BHs are involved in galaxy mergers depends crucially on how many galaxies host BHs and on the galaxy merger history. It is therefore necessary to start with full cosmological models including BH physics and a careful dynamical treatment. After galaxies have merged, however, the BHs still have a long journey until they touch and coalesce. Their dynamical evolution is radically different in gas-rich and gas-poor galaxies, leading to a sort of ``dichotomy" between high-redshift and low-redshift galaxies, and late-type and early-type, typically more massive galaxies. 

\keywords{galaxies: kinematics and dynamics, galaxies: nuclei, black hole physics}
\end{abstract}

\firstsection 
\section{Introduction}\label{sec:intro}
This  contribution is a short update and review on the theoretical aspects of the dynamics of massive black holes (BHs) in galaxy mergers. See \cite{Colpi2011} and \cite{2014SSRv..183..189C} for comprehensive reviews. For reviews and updates on observations, see also Komossa and Greene (both in this volume), and  \cite{2015ASSP...40..103B}.

Massive BHs are routinely found in the centers of most nearby galaxies. They should naturally grow along with their host galaxies through BH-BH mergers \citep{VHM} and accretion of gas, and influence their galaxies through AGN (Active Galactic Nucleus) feedback \citep{DiMatteo2005}. The relative role of BH-BH mergers and gas accretion in growing the population of BHs can be estimated using a simple argument, hinging on Soltan's seminal work \citep{Soltan1982}: the BH mass density increases by accretion by more than one order of magnitude in the last $\sim$10 Gyr \citep{1999MNRAS.303L..34F,YuTremaine2002,Elvis2002,Hop_bol_2007,2015arXiv150504940M}. We expect, therefore, that accretion is responsible for the bulk of BH growth. However,  BH-BH mergers are likely to be important for growing high-mass BHs in gas-poor galaxies \citep{2014MNRAS.440.1590D}.

The main open questions about BHs in galaxy mergers, 35 years after the pioneering paper by \cite{BBR1980} are: Which galaxy mergers lead to BH-BH mergers? For how long BH binaries linger before coalescing via emission of gravitational waves? When and where merging BHs can be detected? In the last few years a sort of ``dichotomy" in BH and galaxy mergers has become apparent. High-redshift and small galaxies tend to be gas-rich: gas will also drive BH dynamics. Low-redshift and large galaxies are gas poor, and BH dynamics will be dominated by interactions with stars. This hints to a different BH dynamical evolution in different classes of galaxies, and at different cosmic times. Intriguingly, these different classes of objects will be probed by different gravitational wave experiments, a large space-based interferometer such as eLISA will probe relatively low-mass BHs ($\lesssim 10^7 \msun$) up to a high redshift, while the Pulsar Timing Arrays can probe high mass BHs ($\gtrsim 10^8 \msun$) at low redshifts ($z<2$).

\section{A massive BH binary's journey}\label{sec:journey}

Studying BH mergers is a severely multi-scale problem: the merger rate of halos and galaxies, hosting BHs in their centers, is driven by the cosmic web on large scales (tens to hundreds of Mpc),  but the final merger of BHs, their plunge into each other's event horizon, occurs at the Schwarzschild radius, $\sim 10^{-7}$ pc for a $10^6 \msun$ BH. We will briefly summarize the various stages here, from the merger of the halos, to the {\it pairing} of BHs in merging galaxies, until they form a gravitationally bound {\it binary}. The reader should keep in mind that at the current time initial and boundary conditions for most studies are idealized and not self-consistently linked to the previous evolutionary stage, because of the enormous dynamical range involved.

\subsection{Context: the cosmic merger rate of halos and galaxies}
\begin{itemize}
\item Cosmological simulations (Mpc $-$ 100 pc). We need to estimate the merger rate of halos with mass from $10^6 \msun$  at  $z\sim 20-30$, where BH formation occurs, to $10^{15} \msun$ at redshift $z\sim0$.  This is a real challenge: to have a statistical sample one needs uniform volume cosmological simulations covering hundreds of Mpc$^3$: this limits the minimum halo mass that can be resolved to $\sim 10^9\msun$, and the spatial resolution to, at best, hundreds of pc, so the information on the evolution of the BHs, and the small-scale dynamics are lost. To overcome this limit and obtain realistic orbital decay, dynamical friction, the force exerted by the gravitational wake caused by a massive object, a BH in our case, moving in an extended medium, must be added as a sub-grid correction \citep{2012MNRAS.420.2662D,2015MNRAS.451.1868T}.  Increasing the dark matter mass resolution is also necessary, to avoid numerical noise \citep{Bellovary10,2015MNRAS.451.1868T}

\item Semi-analytical models (Mpc $-$ sub pc).  Alternatively, one can use semi-analytical models \citep[e.g.,][]{VHM,Sesana2011,2012MNRAS.423.2533B}, which can probe an equivalent volume 100 Gpc$^3$ with minimum halo mass of  $10^5-10^6 \msun$. The advantage of an analytic approach is that in principle it has unlimited spatial resolution, but at the cost of not fully capturing non-linear processes that cannot be described by well behaved mathematical functions (e.g., galaxy mergers). Encouragingly however, the merger rates determined by both techniques are similar in the portions of the parameter space where comparisons can be made. 
\end{itemize}

\subsection{Gas-driven dynamics:}
\begin{itemize}
\item Pairing: galaxy merger simulations (100 kpc $-$1pc) start from idealized initial and boundary conditions, thus loosing information about the cosmic web (e.g., how cosmic flows replenish gas in galaxy discs). The highest achieved numerical resolution is $\sim$1-10 pc (when gas and star formation are included), and dynamical friction is well resolved \citep{2007Sci...316.1874M,Callegari2009,Callegari2011,Svanwas2012,2015MNRAS.447.2123C,2015MNRAS.449..494R}.  In most cases when the mass ratio of the merging galaxies is $q >$0.1, the two BHs form a gravitationally bound binary in the end, especially when a large bound stellar nucleus speeds up BH pairing \citep{Yu2002,VW2014}. In gas-rich environments, at the end of the simulations the BHs are embedded in  nuclear discs. 

\begin{figure}[tbh]
\vspace*{-0.0 cm}
\centering
 \includegraphics[width=0.49\textwidth]{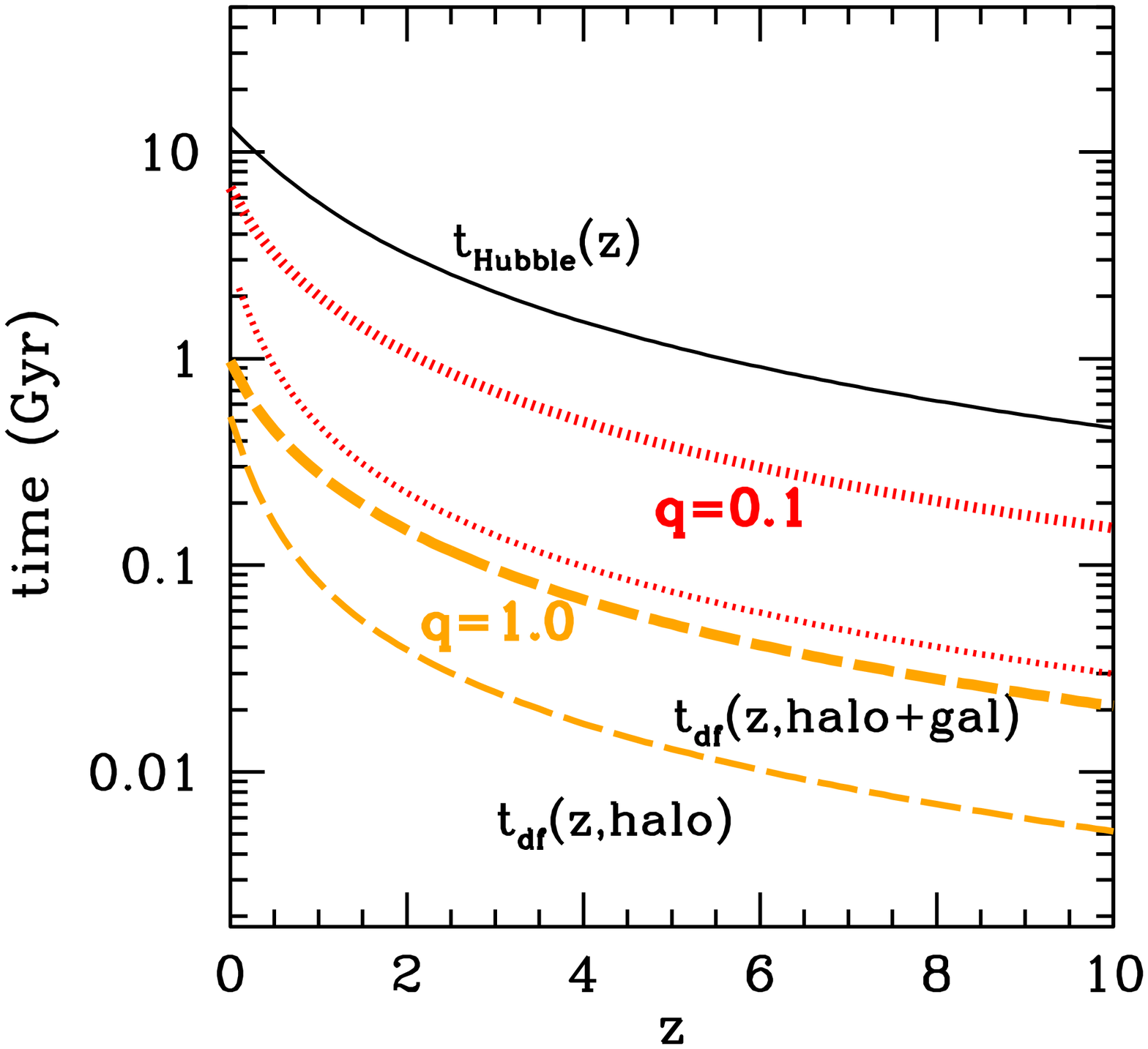}\hfill
 \includegraphics[width=0.49\textwidth]{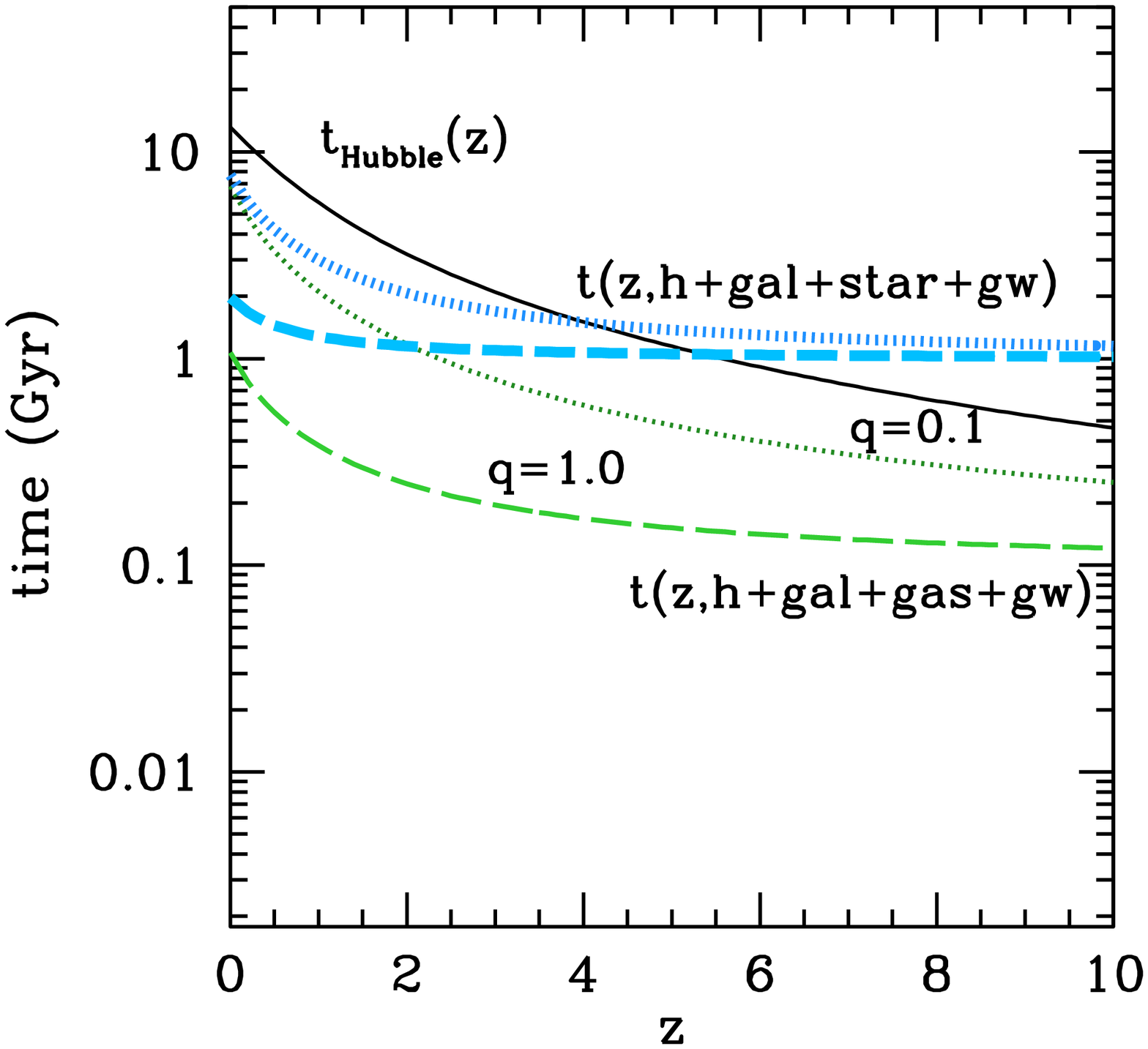} 
\vspace*{-0.5 cm}
 \caption{Left: Merging time for halos only (thin curves), or halos+galaxies (thick curves) for mass ratios of 0.1 (dotted) and 1 (dashed), averaging over orbital parameters. The whole process can take up to several Gyr at $z<1$. Right:  time from when the encounter between halos takes place, all the way to the coalescence of a BH binary (equal-mass binary of $10^8 \msun$ with zero eccentricity), assuming a gas-dominated (thin green curves) or stellar-dominated (thick blue curves) environment. Comparison of the left and right panels allows the reader to appreciate where the bottlenecks lie at different cosmic times and mass ratios. Note that we do not expect many mergers to be stellar dominated at high-redshift (conservatively, $z>1$), when most galaxies are gas-rich.}
   \label{fig:time1}
\end{figure}

\item Binary: nuclear discs (kpc $-$ 0.1 pc). These studies start from idealized initial conditions \citep{Escala2005,Dotti2007,Dotti2009,2013ApJ...777L..14F,2013ApJ...764...14A,2014ApJ...780...84D,2015MNRAS.453.3437L}, and the dynamics of the binary sensitively depends on thermodynamical properties of the gas disc (i.e., hot, cold, lumpy, star formation, SN feedback). The effect of the AGN feedback is usually neglected. The BH separation if found to reach a resolution limit, i.e., sub-pc scales, within 1-100 Myr. They are then expected to evolve through interaction with the accretion disc(s) surrounding them, through the so called circumbinary disc phase.

\item Binary: circumbinary discs (0.1 $-$ 0.001 pc). A binary clears a cavity in its surroundings due to the binary's tidal torques. This was feared to be a show-stopper for further orbital decay, but recent simulations indicate that despite strong binary torques, accretion into the central cavity continues unhindered and may even be enhanced relative to the single BH case. If so, this has two important implications: (1)  sufficient conditions exist for efficient transport of angular momentum through the circumbinary disc and therefore migration to the gravitational wave-dominated regime ($\sim 0.01$ pc) should occur rapidly, within$\sim$1-10 Myr \citep[e.g.,][]{ArmitageNarajan2005,2008ApJ...672...83M, 2012A&A...545A.127R,2012ApJ...749..118S,2012ApJ...755...51N,2013MNRAS.436.2997D,2015MNRAS.447L..80F}, but also (2)  accretion powered binary BHs can shine as AGNs. 
\end{itemize}

\begin{figure}[tbh]
\vspace*{-0.0 cm}
\centering
 \includegraphics[width=0.49\textwidth]{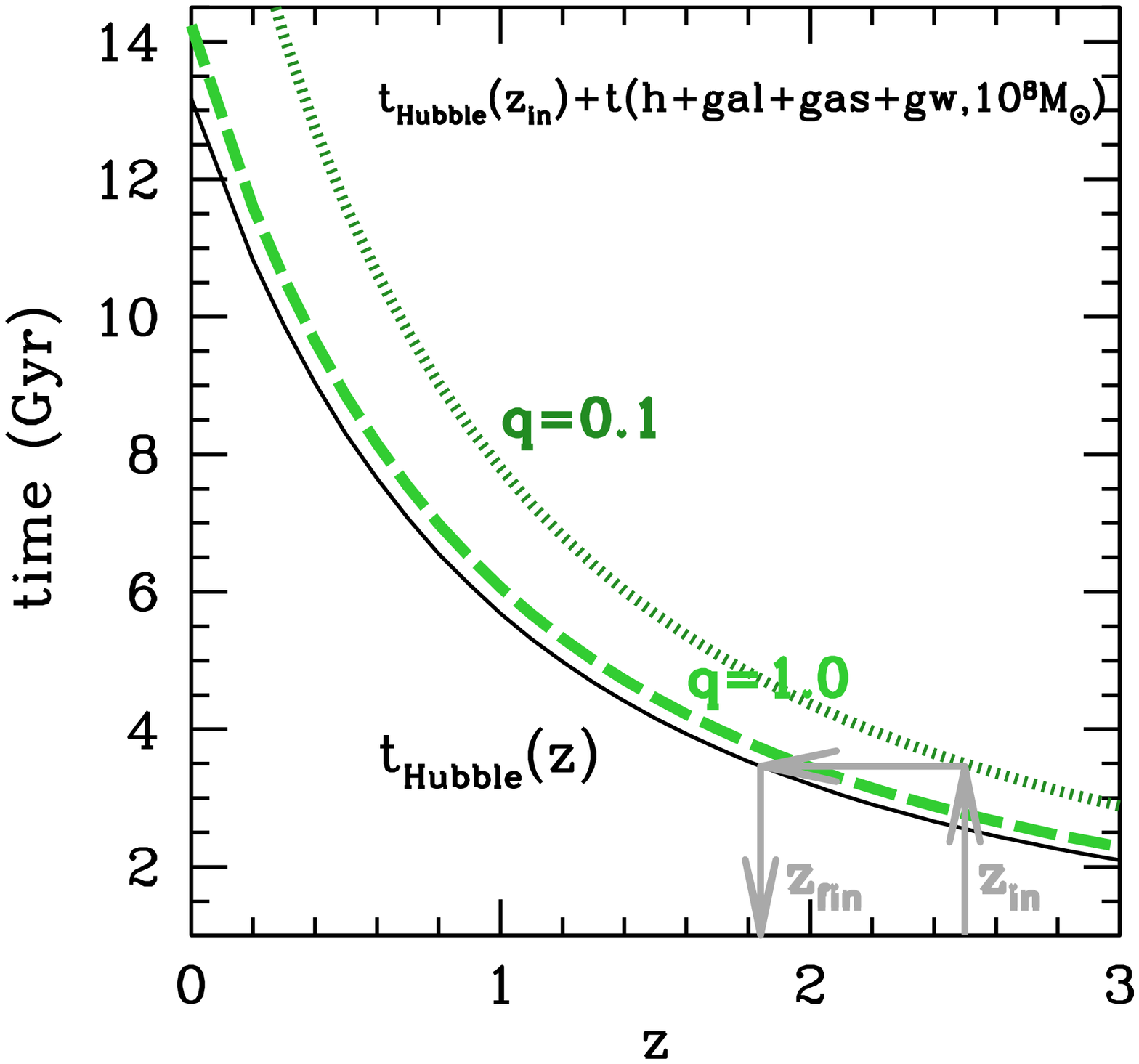}\hfill
 \includegraphics[width=0.49\textwidth]{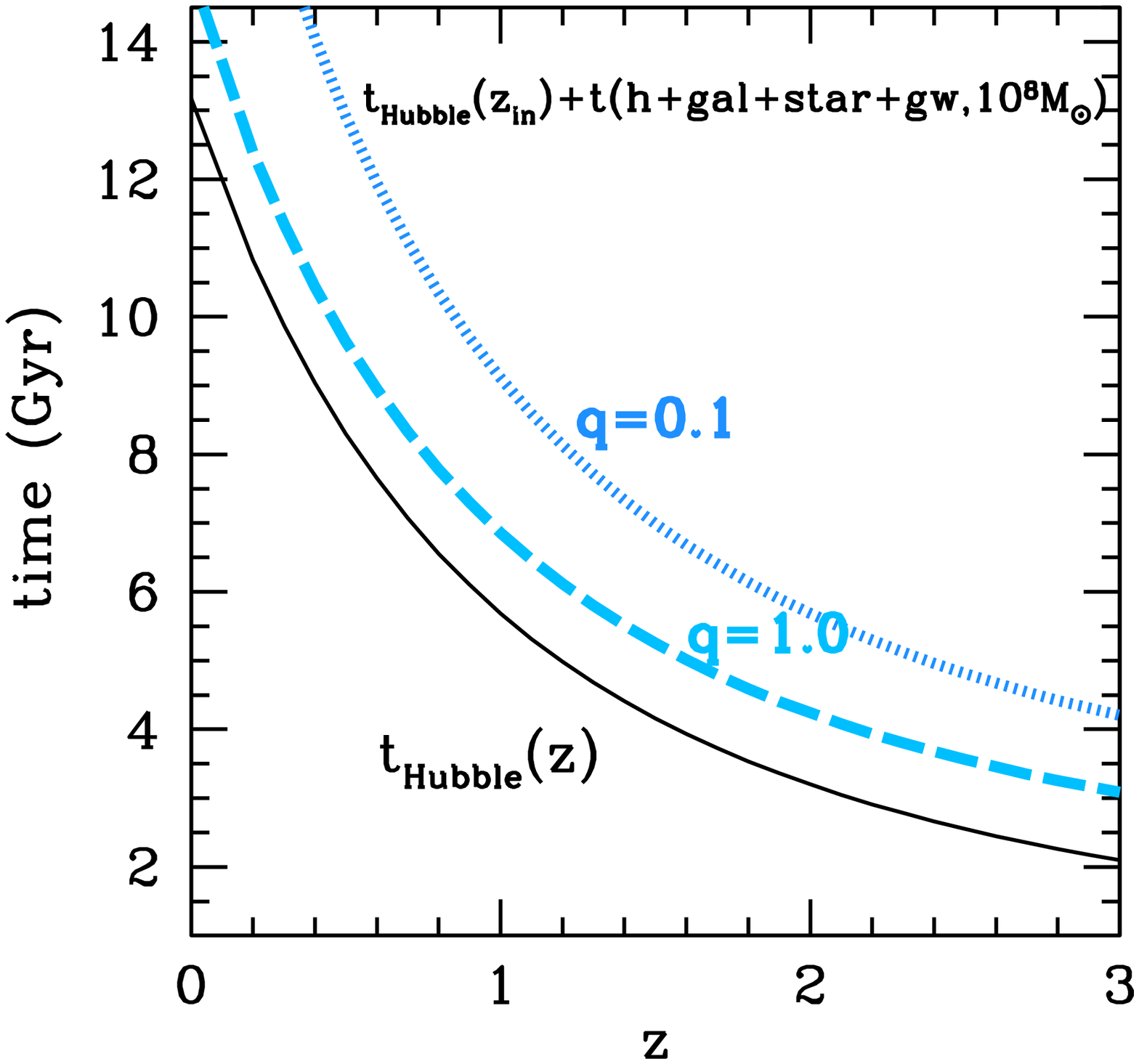} 
 \vspace*{-0.5 cm}
 \caption{Left: Cosmic time at which a binary of $10^8 \msun$ with zero eccentricity in mergers involving halos and galaxies with mass ratio 0.1 and 1 which started at $z=z_{\rm in}$ will merge, assuming that the dynamical evolution is {\it gas-dominated}. To find the redshift $z_{\rm fin}$ at which the merger takes place, do as follows. Starting from a given initial redshift $z=z_{\rm in}$, draw a vertical line. When it intersects the curve of interest, draw a horizontal line until the Hubble time (black) is intersected. Draw a vertical line: the corresponding redshift marks when the binary, which started at $z=z_{\rm in}>z_{\rm fin}$, has merged. E.g., if you want to know when a merger with mass ratio $\sim 0.1$ which started at $z_{\rm in}=2.5$ will complete, follow the grey arrows, to find that the total merger time is $\sim 3.5$ Gyr, and coalescence happens at $z_{\rm fin}\sim 1.84$. Right: same for  {\it stellar-dominated} dynamics.}
   \label{fig:time2}
\end{figure}

\subsection{Stellar-driven dynamics:}

\begin{itemize}
\item Direct N-body simulations (kpc $-$ 0.01 pc). Simulations using only collisionless particles start from idealized initial conditions, and well within the galaxy merger phase when BH separation is only few kpc (compared to 100 kpc for standard galaxy merger simulations). Dynamical friction and scattering between BHs and stars are well resolved \citep[e.g.,][]{Gualandris,2012ApJ...749..147K,2014ApJ...785..163V,2015ApJ...810...49V}, and show that when the separation reaches $\sim$pc scale, 3-body scattering between the binary and low angular momentum stars dominate the orbital decay.
\item Monte Carlo methods/ scattering experiments (0.01$-$ kpc) start from idealized initial conditions, and normally study only the gravitationally bound binary phase, but are very flexible and efficient. Coupling N-body results with those of these codes has shown that the Òlast parsec problemÓ, i.e. running out of low-angular momentum stars (Begelman, Blandford \& Rees 1980) is not a ``problem".  The evolution of BH binaries continues at nearly a constant rate leading to merger in less than $\sim$1 Gyr \citep{2006ApJ...642L..21B,2015ApJ...810...49V,2015arXiv150506203H,2015arXiv150502062S}.
\end{itemize}

\subsection{Merger proper} 

\begin{itemize}
\item Numerical Relativity, analytical techniques ($<0.001$ pc). The merger time depends on the mass, eccentricity, spin and spin configuration of the binary and sensibly on the separation \citep[e.g.,][]{Peters1964,Campanelli2009}.  As soon as the binary reaches the domain of gravitational wave driven inspiral, the merger speeds up rapidly. The gravitational wave signal becomes
detectable months prior to merger and the merger proper lasts several minutes \citep{Pau2012}. \\
\end{itemize}

How long does this all take? We have to consider that first halos need time to merge \citep{2008MNRAS.383...93B}, then it's the turn of galaxies \citep{2014ApJ...789..156M}, which brings us to the end of the pairing phase, so to speak. The total timescales (halo+galaxy) for some typical mergers, averaged over orbital parameters, are shown in Fig.~\ref{fig:time1} (left).  Then the gravitationally bound binary phase starts, and depending on the environment (gas or stars), merger timescales differ. In Fig.~\ref{fig:time1} (right) we show the total merger time, from when the galactic halos touch to when the BHs coalesce, for some simplified cases. Fig.~\ref{fig:time2} shows the same results in a different way, allowing one to estimate when a binary will merge, given an initial redshift when the halo merger begins. We assumed 100 Myr from binary formation to coalescence for the gas-driven case and the fit proposed by \cite{2015arXiv150502062S} for the stellar-driven case. The key uncertainty in the gas-driven scenario is whether at $z<2$ there is enough gas to drive large binaries to merge (see Dotti et al. 2015).  When conditions for merger are fulfilled it is possible to calculate for example that a circular, equal mass binary with mass of $10^8 \msun$ will coalesce by $z=0$ if the halo merger started by $z\sim0.1-0.2$.  A $q=0.1$ binary will coalesce by $z=0$ if the halo merger started by $z\sim0.4-0.5$. 


\begin{table}
  \begin{center}
  \caption{Dual fraction in a 1:2 spiral-spiral merger at $L_{bol}>10^{44}$ erg/s}
  \label{tab:over}
 {\scriptsize
  \begin{tabular}{|l|c|c|c|}\hline 
 & {\bf Dual time (Myr)} & {\bf Dual AGN fraction} & {\bf Observation Type} \\ \hline
No cutoff 		& 12      	& 19.2\%             &  \\
$d>$1 kpc   	& 10         	& 16.5\%             & Imaging (HST) \\
$d>$10 kpc   	& 0.06       	& 0.1\%            	 & Imaging (SDSS) \\
$v>150$ km/s  	& 3           	& 4.8\%              	 & Spectroscopy \\ 
\hline
  \end{tabular}
  }
 \end{center}
\vspace{1mm}
\end{table}

From Fig.~\ref{fig:time1} we can estimate where bottlenecks lie. For the gas-dominated case, at $z\gtrsim$2 the  circumnuclear/circumbinary disc phases are the longest, BHs are therefore expected to linger at $\sim$ pc separations, what are typically referred to as `binary BHs'. At $z\lesssim 2$, in principle,  the dynamical friction phase (left panel) is longer than 0.1~Gyr, the timescale  we assumed here for the circumnuclear/circumbinary disc phases, therefore we should expect mostly `dual BHs' at $\sim$ kpc separations. However, \cite{2015MNRAS.448.3603D} find that  the amount of gas available in the circumbinary disc phase is not sufficient for shrinking the binary effectively. If this is taken into account, a population of binary BHs, at $\lesssim$ pc separations is also to be expected. For the star-dominated case, at mass ratios q$\sim$1 stellar dynamical friction and scattering phases are equally long, implying existence of both dual and binary BHs. For mass ratios q$\sim$0.1 the dynamical friction phase is the longest, therefore we should look for dual BHs. In all cases, if enough gas is present near the BHs, we can hope to detect them as dual or binary AGN.

\section{How many dual/binary BHs can we observe?}\label{sec:pred}
Searches for dual AGN ($\sim$ kpc separation) are ongoing. Current estimates suggest that dual AGN are $\sim$1\% of the AGN population (see Greene's contribution in this volume). We can use simulations of galaxy mergers to estimate the time during a merger over which dual AGN can be detectable, taking observational limitations which reduce detectable dual emission into account \citep{Svanwas2012,2013MNRAS.429.2594B}. In the first place, the AGN must be sufficiently luminous to the be detected as such, photometrically or spectroscopically. Additionally, in searches that look for double AGN in imaging observations, the AGN should be separated by more than the minimum angular resolution of the instrument. For searches based on Doppler-shifted lines (double-peaked narrow line regions) the velocity separation should be much larger than the spectral resolution, and  gas kinematics can also create double-peaked features in the absence of a dual system.  Table~\ref{tab:over} shows an example from \cite{Svanwas2012}. The total simulated time is 1.35 Gyr, but  BHs shine {\it simultanously} above $L_{bol}>10^{44}$ erg/s for only 12 Myr, therefore, the fraction of all galaxies (active and inactive) hosting a dual AGN is $<1\%$. However, when we look at the AGN population, we have to include only the time that one or more BHs are active at the given threshold. Therefore the {\it total} fraction of dual AGN in the AGN population is $\sim 19\%$. When including the additional constraints (such as spatial and spectroscopic resolution) the fraction of {\it detectable} dual AGN in the AGN population drops to a few percent or less. Notably, accretion operates differently on the two BHs, and both theory \citep{Svanwas2012,2015MNRAS.447.2123C} and observations \citep{2015ApJ...806..219C} find that the secondary, smaller BH, tends to shine at a lower luminosity even though it accretes at a higher Eddington ratio. 

Binary BHs and AGN  ($<$ pc separation) have also proven elusive. \cite{Michiganders} estimated the rate of  sub-parsec binary BHs using semi-analytical models. Detectable sub-parsec binaries prove to be  intrinsically rare, due to a combination of effects. On one hand, the merger rate of BHs increases with decreasing mass. On the other hand, the lifetime of  binaries decreases with decreasing mass, making their detection hard \citep{2009ApJ...700.1952H}. Detectable binaries represent a fraction $\sim 0.1\%$  of unabsorbed quasars (M$_i<$-22) in their theoretical sample at $z<0.7$. The short lifetime of low-mass binary BHs suggests that looking for faint sources should not lead to a significant increase in the number of detectable sub-parsec binary quasars.  However, the number of observable binary systems should increase rapidly with redshift.

\section{Conclusions}\label{sec:conclude}

BHs in merging galaxies have a long journey. Beginning to end, it takes between 1 and 10 Gyr. Most BH binaries, however, should merge by $z=0$. The main caveat in  theoretical studies and predictions is that this is a severely multi-scale problem, and most studies are, consequently, highly idealized and not connected self-consistently to the previous ``level" of simulations or theoretical calculations.

Because of lifetimes/observability requirement the fraction of detectable dual and binary BHs is expected to be low. Although a variety of signatures have been predicted by theoretical studies, in practice, only a few approaches have been used to systematically search for binaries in observational campaigns. Future interactions between theory and observations will hopefully bring us a wealth of new information and everyone lived happily ever after.

\section{Acknowledgments}

MV acknowledges funding support for this research from SAO, through Award Number TM1-12007X, from NSF, through Award Number AST 1107675, and from a Marie Curie FP7-Reintegration-Grants within the 7th European Community Framework Programme (PCIG10-GA-2011-303609). TB  acknowledges support from the National Science Foundation under grant No. AST-1211756 and support from the National Aeronautics and Space Administration under grant No. NNX15AK84G issued through the Astrophysics Theory Program. 

\bibliographystyle{mn2e}
\bibliography{focus_meeting_volonterietal-1}

%
%
%

%


\end{document}